\documentclass{article}

\usepackage{amsmath, amsthm, mathtools, amssymb, color, float, eqnarray, booktabs, makecell, soul, mdframed}
\usepackage[hyphens]{url}
\usepackage[square,numbers]{natbib}
\usepackage[justification=centering]{caption}
\usepackage{todonotes}
\usepackage{comment}
\usepackage{bm}
\usepackage{hyperref}
\usepackage{adjustbox}
\usepackage{subcaption} 

\usepackage{epstopdf}
\pdfoutput=1

\usepackage[most]{tcolorbox}

\usepackage{graphicx, bm, amsmath, siunitx, longtable,tabularx, float, mathtools, booktabs, eqnarray, color, amsfonts, amssymb, appendix}

\usepackage{amsmath,amssymb,mathtools,eqnarray}  
\usepackage{empheq} 
\usepackage{upgreek}  
\usepackage{systeme}

\newcommand{\T}{^{\mbox{\tiny T}}}

\begin{document}


\title{
Physics-Informed Extreme Theory of Functional Connections Applied to Data-Driven Parameters Discovery of Epidemiological Compartmental Models\\
}
\date{}
\author{Enrico Schiassi, Andrea D'Ambrosio, Mario De Florio, \\
Roberto Furfaro, Fabio Curti}

\maketitle
\begin{abstract}

In this work we apply a novel, accurate, fast, and robust physics-informed neural network framework for \textit{data-driven parameters discovery} of problems modeled via parametric ordinary differential equations (ODEs) called the \textit{Extreme Theory of Functional Connections} (X-TFC). The proposed method merges two recently developed frameworks for solving problems involving parametric DEs, 1) the \textit{Theory of Functional Connections} (TFC) and 2) the \textit{Physics-Informed Neural Networks} (PINN). In particular, this work focuses on the capability of X-TFC in solving inverse problems to estimate the parameters governing the epidemiological compartmental models via a deterministic approach. The epidemiological compartmental models treated in this work are Susceptible-Infectious-Recovered (SIR), Susceptible-Exposed-Infectious-Recovered (SEIR), and Susceptible-Exposed-Infectious-Recovered-Susceptible (SEIR). The results show the low computational times, the high accuracy and effectiveness of the X-TFC method in performing data-driven parameters discovery of systems modeled via parametric ODEs using unperturbed and perturbed data.

\end{abstract}
\section{Introduction}

In the last century, the concern of diseases spread has been in the spotlight for many researchers. A first categorization between the models can be made for deterministic and stochastic models. \\
Deterministic models are the simplest, with fixed input variables. They are also known as compartmental models, because the individuals in the population are assigned to different subgroups, or compartments, each of which represents a specific condition of the individual in the epidemic situation \cite{breda}. Derivatives in time are used to express the transition rates of individuals from a compartment to another, thus the model is constructed as a system of ordinary differential equations (ODEs). \\
Stochastic models take into account variations in input variables, and provide results in terms of probability. A stochastic model, unlike a deterministic one, allows random variations in one or more inputs over time, therefore an estimate of the probability distributions of the outcomes can be made. Specifically, the variables changing in time can be the exposure risk, recovery rate, and other disease dynamics. Being able to insert the variability of the input data, the stochastic models have a more complex structure than the deterministic ones, but manage to be more adherent to reality \cite{britton}.\\
A second categorization between the models can be made by taking or not taking into account the vital dynamics. The vital dynamics represents the demography dynamics, that is the case in which the naturally occurring births and deaths are included \cite{hethcote}.\\
In this paper, deterministic models with vital dynamics are studied. Precisely, the Susceptible--Infectious--Recovered (SIR), Susceptible--Exposed--Infectious--Recovered (SEIR), and Susceptible--Exposed--Infectious--Recovered--Susceptible (SEIR) models are considered, with the vaccination factor for one of those \cite{huang,trawicki}.\\
The aim of this work is to estimate various epidemiological model parameters by using the newly developed \textit{Extreme Theory of Functional Connections} (X-TFC) \cite{etfc}. This method aims to solve forward problems, and inverse problems (\textit{data-driven parameters discovery}) involving parametric DEs, in different perturbation scenarios.\\
To solve mathematical and physical inverse problems there are basically two main approaches: deterministic and probabilistic. The deterministic approach tackles inverse problems using standard optimization techniques.
According to these techniques a set of optimal parameters is found, which minimizes the difference between simulated and real data. 
However, inverse problems are known to be ill-posed  \cite{kimes2000} and hence, it becomes very hard to determine the uncertainty in the retrieved quantities mainly due to the noise in the observed data and the uncertainty in the real values of the input parameter that are not tuned.\\
As stated in \cite{bp},  inverse problems to parameters' estimation are in general ill-posed because, 1) the problem is non-unique because the higher number of unknowns than data/measurements, and 2) the stability of the solution, to noise in the data and modeling errors, is generally not guaranteed.
Standard optimization techniques consider the tuning quantities to be deterministic. Therefore, the inverse problems' outputs are fixed quantities.
However, these quantities are affected by uncertainties that need to be estimated. The issue is that uncertainty quantification (usually done via regularization techniques) is not trivial to perform; and it can lead to poor results, in particular when the problem is ill-posed. Moreover, nonlinear or non-convex inverse problems have more local minimum solutions. Thus, more than one acceptable solutions can be computed, and it becomes challenging to select the best one via the classical optimization framework \cite{bc11}.\\
To overcome this issue the probabilistic approach can be used, in particular Bayesian inversion techniques. 
In the Bayesian inversion framework, the quantities to be estimated are considered as random variables. Thus, the output of the inverse modeling is the probability distribution for each of those parameters. 
Therefore, with the probabilistic approach, the degree of uncertainty of the values of the quantities to be retrieved is included in their probability distributions \cite{glam}.\\
Nevertheless, in this work, we tackle the inverse problem for data-driven parameters discovery of epidemiological models via a deterministic approach. We show that solving these problems via Physics-Informed Neural Network (PINN) methods, such as X-TFC, mitigates the ill-posedness of the inverse problems toward modeling errors and noise in the data. This is due to the fact that the physics of the problem, modeled via a parametric DE, acts as a regulator during the search of the optimal parameters (i.e. the NN training).\\
\color{black}
In this manuscript, in section \ref{sec: method} the X-TFC framework is explained, along with the step-by-step derivation of the constrained expression used for initial value ODEs, and the ELM algorithm. In section \ref{sec: results} the application of the X-TFC for data-driven discovery of the parameters governing a few of the most common epidemiological compartmental models is presented. Finally in section \ref{sec: results} the results are presented and discussed. 
\section{Extreme Theory of Functional Connections}\label{sec: method}
The Extreme Theory of Functional Connections framework can be used for solving forward and inverse problems involving parametric DEs with high precision and low computational time. 
The method for solving direct problems involving parametric DEs, is introduced and presented by Schiassi et al. in \cite{etfc}. 
When data is needed to solve the equation with high accuracy, the solution of the parametric DE is called a \textit{data-driven solution} \cite{raissi}. When tackling inverse problems involving parametric DEs, the parameters governing the DE are unknown and, thus, they need to be quantified. These kind of problems are called \textit{data-driven parameters discovery} of parametric DEs \cite{raissi}, as the goal is to identify the parameters that govern the equation by comparison of the equation solution with data. For instance, a typical field where solving inverse problems is of interest is the remote sensing \cite{glam,schiassi,hapke81,hapke96}. For instance, in Ref. \cite{hapke2002}, the authors combine radiative and heat transfer equations to create a set of parametric DEs. The solutions of this system of equations is compared with real data to retrieve the the grain size and the thermal inertia of planetary regoliths, which are the parameters governing the physics of the problem.\\
As previously stated, the focus of this work is to apply the X-TFC for data-driven parameters discovery of compartmental epidemiological models such as SIR, SEIR, and SEIRS. In the remain of this section we will explain how the X-TFC is applied to tackle this kind of problems. 
Such models are systems of ODEs, where the constraints are on the initial values of the solutions of these systems. That is, these problems are initial value problems (IVPs). Therefore, in this section we will also present the step-by-step derivation of the constrained expressions for these types of problems. Finally we will briefly explain how the ELM works and provide some theoretical guarantees for the convergence of this learning algorithm that are formally presented and proved in \cite{ELM}.

\subsection{Method}

\begin{figure}[H]
    \centering\includegraphics[width=1\linewidth]{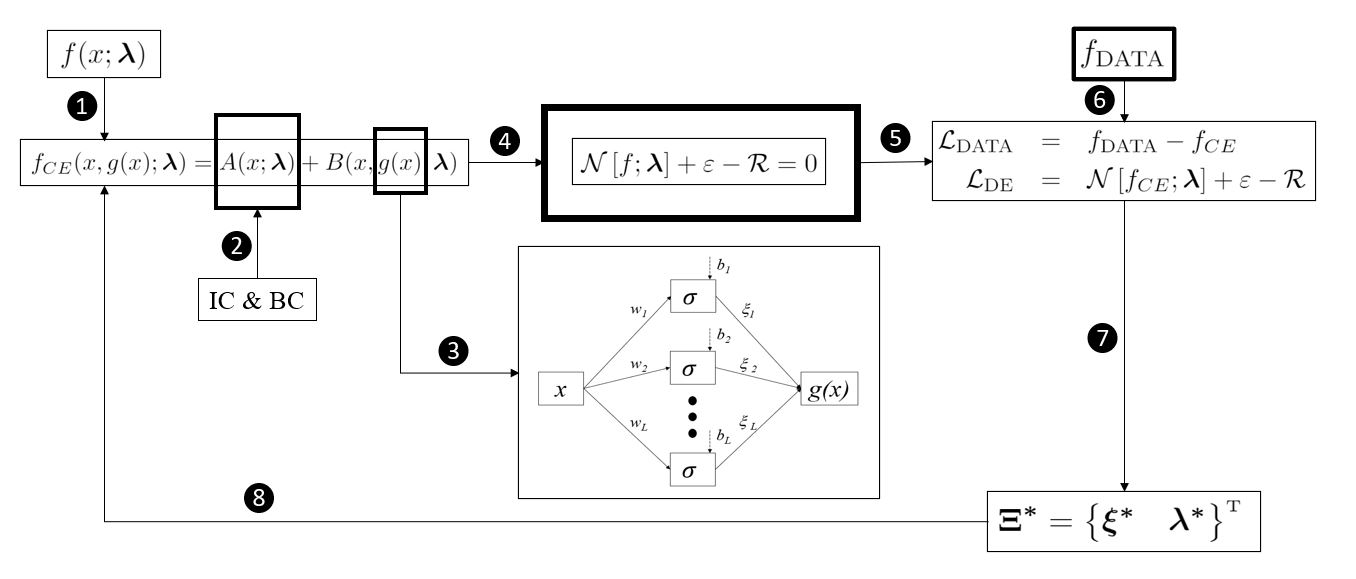}
    \caption{Schematic of the X-TFC framework for solving inverse problems:\\
    (1) Approximate the latent solution(s) with the CE;\\
    (2) Analytically satisfy the ICs/BCs;\\
    (3) Expand with the single layer NN (trained via ELM);\\
    (4) Substitute into the DE (that can be also a system of DEs);\\
    (5) Build the DE Losses (that drive the training of the network informing it with the physics of the problem);\\
    (6) Build the data Losses (the data can be provided on the solutions and/or on their derivatives);\\
    (7) Train the network;\\
    (8) Build the approximate solution (with the estimated optimal parameters).}
    \label{covid_schematic}
\end{figure}

In this work, we will focus on systems of ODEs (SODEs) used to describe epidemiological compartmental models.  
In general, we can express parametric ODEs, in their implicit form as,
\begin{equation}\label{eq:parDEgen}
    \mathcal{N}\left[f;\bm{\lambda}\right] +\varepsilon - \mathcal{R} = 0
\end{equation}
subject to constraints given by the boundary conditions (BC) and initial conditions (IC). In Equation \eqref{eq:parDEgen}, 
 $f:=f(x;\bm{\lambda}(x))$ is the unknown (or latent) solution ,
 $ x \in \mathbb{D} \subseteq \mathbb{R}$,
$\bm{\lambda}:=\bm{\lambda}(x) \in \mathbb{L} \subseteq \mathbb{R}^m $ are the parameters governing the parametric ODE 
\footnote{
In general, even if it is not reported in the notation,
$f$ is a function of $x$, and it is parameterized by $\bm{\lambda}$, that in general can be $x$ dependent as well.
},
$\mathcal{N}\left[ \cdot ; \bm{\lambda} \right]$ is a linear or non-linear operator acting on $f$ and parameterized by $\bm{\lambda}$,
$\varepsilon$ is the modeling error that is negligible when solving \textit{exact problems},
and $\mathcal{R}$ is a known term that in general can be $x$ dependent and parametrized by $\bm{\lambda}$ as well.\\
The first step in our physics-informed NN framework is to approximate the latent solution $f$ with a constrained expression,
   \begin{equation}
        f(x; \bm{ \lambda }) = f_{CE}(x, g(x); \bm{\lambda}) = A(x; \bm{\lambda}) + B(x, g(x); \bm{\lambda}),
    \end{equation}
where 
$A (x; \bm{\lambda})$ analytically satisfies the constraints, and 
$B(x, g (x); \bm{\lambda})$ projects the free-function $g(x)$ onto the space of functions that vanish at the constraints \cite{leake}. 
According to the X-TFC method we chose the free-function, $g(x)$, to be a single layer NN, trained via ELM \cite{ELM}. That is,
\begin{equation}
g(x) = \sum_{j=1}^{L} \xi_j\sigma \left(w_jx + b_j \right):= [\sigma_1,...,\sigma_L] \, \bm{\xi} := \bm{h}\T \bm{\xi}, 
\end{equation}
where $L$ is the number of hidden neurons, $w_j \in \mathbb{R}$ is the input weights vector connecting the $j^{th}$ hidden neuron and the input nodes, 
$\xi_j \in \mathbb{R}$ with $j=1,...,L$  is the $j^{th}$ output weight connecting the $j^{th}$ hidden neuron and the output node, and $b_j$ is the bias of the $j^{th}$ hidden neuron,
$\sigma(\cdot)$ are activation functions, and $\bm{h} = [\sigma_1,...,\sigma_L]\T$.
According to the ELM algorithm \cite{ELM}, biases and input weights are randomly selected and not tuned during the training, thus they are known hyperparameters. The activation functions, $\sigma(\cdot)$, are also known, as they are user selected. Thus, the only unknowns NN hyperparameters to compute are the output weights $\boldsymbol{\xi} = \left[ \xi_1,...,\xi_L \right]\T$. Hence we can write,
\begin{equation*}
    f(x; \bm{\lambda}) = f_{CE}(x, g(x); \bm{\lambda}) = f_{CE}(x,\bm{\xi}; \bm{\lambda}).
\end{equation*}
The step-by-step process to derive the constrained expression is provided in Section \ref{sec:CE}. Once $f$ is approximated with a NN, the second step of the X-TFC physics-informed method is to define the loss functions,
\begin{eqnarray}
    \mathcal{L}_{\text{DATA}} &=&  f_{\text{DATA}} - f_{CE} \\
    \mathcal{L}_{\text{DE}} &=&  \mathcal{N}\left[f_{CE};\bm{\lambda}\right] +\varepsilon -\mathcal{R}
\end{eqnarray}
where $f_{\text{DATA}}$ are the real data, that eventually can be perturbed. Once the losses are defined, we need to defined the vectors with all the unknowns, that are the $\bm{\xi}$ coefficients and the parameters governing the equations $\bm{\lambda}$,
\begin{equation*}
    \bm{\Xi} = \begin{Bmatrix} \bm\xi & \bm{\lambda} \end{Bmatrix}\T
\end{equation*}
Now, by combining the losses, an augmented loss function vector is formed as follows,
\begin{equation}\label{gen_aug_loss}
    \mathbb{L} = \begin{Bmatrix} 
    \mathcal{L}_{\text{DATA}}, &  \mathcal{L}_{\text{DE}}  \end{Bmatrix} \T
\end{equation}
and enforcing that for a true solution, this vector should be equal to $\bm{0}$. This allows the unknowns to be solved via different optimization schemes, e.g., least-square for linear problems \cite{LDE} and iterative least-squares for non-linear problems \cite{NDE}. 
When solving inverse problems for parameter estimation, the iterative least-square method is required. 
Thus, the estimation of the unknowns are updated at each iteration as follows,
\begin{equation}\label{eq:gen_xi_update}
    \bm{\Xi}_{k+1} = \bm{\Xi}_k - \Delta \bm{\Xi}_k
\end{equation}
where the $k$ subscript refers to the current iteration.
In general, the $\Delta \bm{\Xi}_k$ term can be defined by performing classic linear least-square at each iteration of the iterative least-square procedure as follows,
\begin{equation}\label{eq:gen_xi_delta}
    \Delta \bm{\Xi}_k = \Big(\mathbb{J}(\bm{\Xi}_k)\T \mathbb{J}(\bm{\Xi}_k) \Big)^{-1} \mathbb{J}(\bm{\Xi}_k)\T \mathbb{L}(\bm{\Xi}_k)
\end{equation}
where $\mathbb{J}$ is the Jacobian matrix containing the derivatives of the losses with respect to all the unknowns. One can consider to compute the Jacobian either by hand or by means of computing tools, such as the Automatic Differentiation of Matlab which is part of the Deep Learning Toolbox since the 2020a version. 
The iterative process is repeated until either of the following conditions are met,

\begin{equation}\label{eq:gen_stop_crit}
    L_2 [\mathbb{L}(\bm{\Xi}_k)] < \epsilon \qquad \text{or} \qquad
    L_2 [\mathbb{L}(\bm{\Xi}_{k+1})] > L_2 [\mathbb{L}(\bm{\Xi}_k)].
\end{equation}
where $\epsilon$ defines some user prescribed tolerance.\\
In figure \ref{covid_schematic}, a schematic that summarize how the X-TFC algorithm works to solving inverse problems is shown.

\subsection{Constrained Expression Derivation}\label{sec:CE}
Since this paper focuses on IVPs, for the convenience of the reader, we will present the step-by-step derivation of the constrained expression for these kind of problems. The interested reader can find the general derivation for a $n + 1$ dimensional constrained expression either in \cite{leake} or \cite{etfc}.
Given a parametric ODE where we have a constraint on the initial value of the solution (i.e. $f(0)= f_0$), the constrained expression for $f$ is the following \cite{TFC},
\begin{equation} \label{eq:ce}
    f_{CE}(x) = g(x) + \eta = \bm{h}\T \xi + \eta\
\end{equation}
By imposing the constraint $f_0$ into Eq. \eqref{eq:ce} we get the following,
\begin{equation} \label{eq:eta}
    \eta = f_0 - g_0 = f_0 - \bm{h}_0\T \xi
\end{equation}
Now by plugging this results back into Eq. \eqref{eq:ce} we get,
\begin{equation} 
    f_{CE} = \big[ \bm{h} - \beta_1\bm{h}_0 \big]\T \xi +\beta_1 f_0,
\end{equation}
where $\beta_1$ is called switching function. For an IVP with one constraint on $f$, we have $\beta_1(z) = \beta_1 = 1$.\\
In general $f$ is defined in $x \in [x_0,x_f]$. Thus we need to map it into the $z$ domain as follows,
\begin{equation}
z = z_0 + c(x - x_0) \quad \leftrightarrow \quad x = x_0 + \frac{1}{c}(z - z_0),
\end{equation}
where the mapping coefficient $c$ is,
\begin{equation}
    c := \frac{\mathrm{d} f}{\mathrm{d} x} = \frac{\Delta z}{x_f - x_0}
\end{equation}
According to the chain rule of the derivative we then have,

\begin{equation}
    \frac{\mathrm{d^n} f}{\mathrm{d} x^n} =c^n  \frac{\mathrm{d^n} f}{\mathrm{d} z^n}
\end{equation}

The interested readers can find detailed information and the proof of convergence of the ELM algorithm in \cite{ELM}.

\section{Epidemiological Models} \label{sec: results}

In this section, the X-TFC formulation for the data-driven parameters discovery of a series of epidemiological compartmental models is explained in details. The presented models are the SIR, SEIR, and SEIRS, taking into account the vital dynamics and the vaccination (for the SEIR model). As already mentioned, the goal is to estimate the parameters of our interest through solving inverse problems via a deterministic approach.\\
Given fixed parameters, by integration, we solve the systems of ODEs to create a synthetic data-set (with and without noise), through which the parameters that govern the physics of the problem can be retrieved. After building the constrained expressions and the loss functions, the Jacobian matrix (the matrix containing the derivatives of the losses with respect to the unknowns) is computed in order to perform the iterative least-squares and compute the unknowns. 

\subsection{SIR Model}

As first problem, we consider the system of differential equations that govern the classic deterministic \textbf{SIR} (\textbf{S}usceptible-\textbf{I}nfectious-\textbf{R}ecovered) compartmental model, in which individuals in the recovered state gain total immunity to the pathogen, with vital dynamics to take in account the births (that can provide an increase in susceptible individuals) and natural death rates. The DEs governing the SIR model are the following:
\begin{equation}
    \begin{aligned}
        \dfrac{dS}{dt} &= \mu N -\beta \dfrac{SI}{N} - \mu S \\
        \dfrac{dI}{dt} &= \beta \dfrac{SI}{N}  - \gamma I - \mu I \\
        \dfrac{dR}{dt} &= \gamma I - \mu R
    \end{aligned} \qquad \qquad \text{subject to} \qquad \qquad
    \begin{aligned}
        S(t_0) &= S_0 \\
        I(t_0) &= I_0 \\
        R(t_0) &= R_0 \\
    \end{aligned}
    \label{sir}
\end{equation}
where $N = S+I+R$ is the total population, $\mu$ is the birth and natural death rate (considered equal to maintain a constant population), $\beta$ is the infectious rate, and $\gamma$ is the recovery rate. An important parameter to consider is the basic reproduction number $\mathcal{R}_0$, which represents the ratio between $\beta$ and $\gamma$. If $\mathcal{R}_0>1$, an outbreak is going to occur.\\ 
According to the TFC framework, the latent solutions are approximated with the constrained expressions. That is,

\begin{equation}
    \begin{aligned}
      S &=\left(\mathbf{h} - \beta_1\mathbf{h_0} \right)\T \boldsymbol{\xi}_1 + \beta_1 S_0 \\
      I &=\left(\mathbf{h} - \beta_1\mathbf{h_0} \right)\T \boldsymbol{\xi}_2 + \beta_1 I_0 \\
      R &=\left(\mathbf{h} - \beta_1\mathbf{h_0} \right)\T \boldsymbol{\xi}_3 + \beta_1 R_0 \\
    \end{aligned}
\end{equation}

The first three loss functions we present take into account the regression over the data. The last three losses drive the training of the NN informing it with the physics governing the problem. The Loss functions are report below:
\begin{equation}
    \begin{aligned}
        \mathcal{L}_1 &= \Tilde{S} - S  \\
        \mathcal{L}_2 &= \Tilde{I} - I  \\
        \mathcal{L}_3 &= \Tilde{R} - R  \\
        \mathcal{L}_4 &= \Dot{S} - \mu N + \beta \dfrac{SI}{N} + \mu S    \\
        \mathcal{L}_5 &= \Dot{I} -  \beta \dfrac{SI}{N}  + \gamma I + \mu I  \\
        \mathcal{L}_6 &=  \Dot{R} - \gamma I + \mu R \\
    \end{aligned}
\end{equation}
To construct the Jacobian matrix $\mathcal{J}$ we need to compute the derivatives of the Losses with respect to the $\boldsymbol{\xi}$ to compute the approximate solutions of the state variables, whereas the other derivatives are essential to estimate the parameters (in this case $\beta$ and $\gamma$) appearing in the system of eqs. \eqref{sir}. The resultant Jacobian matrix has the following form:
\begin{equation}
    \mathcal{J} = 
        \begin{bmatrix}
            \dfrac{\partial \mathcal{L}_1}{\partial \boldsymbol{\xi}_1}   &     \mathbf{0}   &    \mathbf{0}   &    0   &     0  \\    \mathbf{0}    &     \dfrac{\partial \mathcal{L}_2}{\partial \boldsymbol{\xi}_2}   &      \mathbf{0}    &    0    &     0  \\   \mathbf{0}  &    \mathbf{0}   &     \dfrac{\partial \mathcal{L}_3}{\partial \boldsymbol{\xi}_3}   &    0   &     0  \\    \dfrac{\partial \mathcal{L}_4}{\partial \boldsymbol{\xi}_1}   &     \dfrac{\partial \mathcal{L}_4}{\partial \boldsymbol{\xi}_2}   &     \dfrac{\partial \mathcal{L}_4}{\partial \boldsymbol{\xi}_3}   &     \dfrac{\partial \mathcal{L}_4}{\partial \beta}   &     0  \\    \dfrac{\partial \mathcal{L}_5}{\partial \boldsymbol{\xi}_1}   &     \dfrac{\partial \mathcal{L}_5}{\partial \boldsymbol{\xi}_2}   &     \dfrac{\partial \mathcal{L}_5}{\partial \boldsymbol{\xi}_3}   &     \dfrac{\partial \mathcal{L}_5}{\partial \beta}   &     \dfrac{\partial \mathcal{L}_5}{\partial \gamma}    \\
             \mathbf{0}   &     \dfrac{\partial \mathcal{L}_6}{\partial \boldsymbol{\xi}_2}   &     \dfrac{\partial \mathcal{L}_6}{\partial \boldsymbol{\xi}_3}   &     0  &     \dfrac{\partial \mathcal{L}_6}{\partial \gamma} 
        \end{bmatrix}
\end{equation}

\subsection{SEIR Model}


The second problem that we aim to solve is the \textbf{SEIR} (\textbf{S}usceptible-\textbf{E}xposed-\textbf{I}nfectious-\textbf{R}ecovered) compartmental model. This model, compared to the previous one, takes into account the incubation period of a virus, i.e. the time in which a subject comes into contact with the virus, but still does not develop its symptoms. Therefore the subject is infected but is not yet considered among the infectious. Also, a vaccination parameter which moves people from the Susceptible to Recovered directly is added. The following is the ODEs system describing the model:

\begin{equation}
    \begin{aligned}
        \dfrac{dS}{dt} &= \mu (N-S) -\beta \dfrac{SI}{N} - \nu S \\
        \dfrac{dE}{dt} &= \beta \dfrac{SI}{N} - (\mu +  \sigma) E \\
        \dfrac{dI}{dt} &= \sigma E  - \gamma I - \mu I \\
        \dfrac{dR}{dt} &= \gamma I - \mu R + \nu S
    \end{aligned} \qquad \qquad \text{subject to} \qquad \qquad
    \begin{aligned}
        S(t_0) &= S_0 \\
        E(t_0) &= E_0 \\
        I(t_0) &= I_0 \\
        R(t_0) &= R_0 \\
    \end{aligned}
\end{equation}
where $N=S+E+I+R$ is the total population, $\mu$ is the birth and natural death rate (considered equal to maintain a constant population), $\nu$ is the vaccination rate, $\beta$ is the infectious rate, $\sigma$ is the rate at which an Exposed person becomes Infectious, and $\gamma$ is the recovery rate.\\
According to the TFC framework, the latent solutions are approximated with the constrained expressions. That is,

\begin{equation}
    \begin{aligned}
      S &=\left(\mathbf{h} - \beta_1\mathbf{h_0} \right)\T \boldsymbol{\xi}_1 + \beta_1 S_0 \\
      E &=\left(\mathbf{h} - \beta_1\mathbf{h_0} \right)\T \boldsymbol{\xi}_2 + \beta_1 E_0 \\
      I &=\left(\mathbf{h} - \beta_1\mathbf{h_0} \right)\T \boldsymbol{\xi}_3 + \beta_1 I_0 \\
      R &=\left(\mathbf{h} - \beta_1\mathbf{h_0} \right)\T \boldsymbol{\xi}_4 + \beta_1 R_0 \\
    \end{aligned}
\end{equation}
The first four loss functions we present take into account the regression over the data. The last four losses drive the NN informing it with the physics governing the problem. The Loss functions are report below:
\begin{equation}
    \begin{aligned}
        \mathcal{L}_1 &= \Tilde{S} - S  \\
        \mathcal{L}_2 &= \Tilde{E} - E  \\
        \mathcal{L}_3 &= \Tilde{I} - I  \\
        \mathcal{L}_4 &= \Tilde{R} - R  \\
        \mathcal{L}_5 &= \Dot{S} - \mu (N-S) + \beta \dfrac{SI}{N} + \nu S    \\
        \mathcal{L}_6 &= \Dot{E} - \beta \dfrac{SI}{N} + (\mu +  \sigma) E  \\
        \mathcal{L}_7 &= \Dot{I} - \sigma E  + (\gamma  + \mu) I \\
        \mathcal{L}_8 &= \Dot{R} - \gamma I + \mu R - \nu S
    \end{aligned}
\end{equation}
To construct the Jacobian matrix $\mathcal{J}$ we need to compute the derivatives of the Losses in respect of the $\boldsymbol{\xi}$ to compute the approximate solutions of the state variables, whereas the other derivatives are essential to estimates the parameters (in this case $\beta$, $\gamma$, and $\sigma$) appearing in the system of eqs. \eqref{sir}. The resultant Jacobian matrix has the following form:
\begin{equation}
    \mathcal{J} = 
        \begin{bmatrix}
            \dfrac{\partial \mathcal{L}_1}{\partial \boldsymbol{\xi}_1}   &     \mathbf{0}   &    \mathbf{0}   &    \mathbf{0}   &     0  &  0   &   0\\    
            \mathbf{0}    &     \dfrac{\partial \mathcal{L}_2}{\partial \boldsymbol{\xi}_2}   &      \mathbf{0}    &    \mathbf{0}    &     0  &   0   &   0\\   
            \mathbf{0}  &    \mathbf{0}   &     \dfrac{\partial \mathcal{L}_3}{\partial \boldsymbol{\xi}_3}   &    \mathbf{0}   &     0  &   0    &   0 \\    
            \mathbf{0}   &    \mathbf{0}   &     \mathbf{0}   &     \dfrac{\partial \mathcal{L}_4}{\partial \boldsymbol{\xi}_4}   &     0   &   0   &   0   \\    
            \dfrac{\partial \mathcal{L}_5}{\partial \boldsymbol{\xi}_1}   &     \dfrac{\partial \mathcal{L}_5}{\partial \boldsymbol{\xi}_2}   &     \dfrac{\partial \mathcal{L}_5}{\partial \boldsymbol{\xi}_3}   &     \dfrac{\partial \mathcal{L}_5}{\partial \boldsymbol{\xi}_4}   &   \dfrac{\partial \mathcal{L}_5}{\partial \beta}  &  0  & 0 \\
            \dfrac{\partial \mathcal{L}_6}{\partial \boldsymbol{\xi}_1}   &     \dfrac{\partial \mathcal{L}_6}{\partial \boldsymbol{\xi}_2}   &     \dfrac{\partial \mathcal{L}_6}{\partial \boldsymbol{\xi}_3}   &     \dfrac{\partial \mathcal{L}_6}{\partial \boldsymbol{\xi}_4}  &    \dfrac{\partial \mathcal{L}_6}{\partial \beta}  &  0  &   \dfrac{\partial \mathcal{L}_6}{\partial \sigma}  \\
            \mathbf{0}   &     \dfrac{\partial \mathcal{L}_7}{\partial \boldsymbol{\xi}_2}   &     \dfrac{\partial \mathcal{L}_7}{\partial \boldsymbol{\xi}_3}   &     \mathbf{0}  &     0  &  \dfrac{\partial \mathcal{L}_7}{\partial \gamma}   &  \dfrac{\partial \mathcal{L}_7}{\partial \sigma} \\
           \dfrac{\partial \mathcal{L}_8}{\partial \boldsymbol{\xi}_1}   &     \mathbf{0}   &     \dfrac{\partial \mathcal{L}_8}{\partial \boldsymbol{\xi}_3}   &     \dfrac{\partial \mathcal{L}_8}{\partial \boldsymbol{\xi}_4}  &     0  &  \dfrac{\partial \mathcal{L}_8}{\partial \gamma}   &  0
        \end{bmatrix}
\end{equation}

\subsection{SEIRS Model}

The last problem we present here, is the \textbf{SEIRS} (\textbf{S}usceptible-\textbf{E}xposed-\textbf{I}nfectious-\textbf{R}ecovered-\textbf{S}usceptible) compartmental model. This model is used in the case when the immunity of recovered individuals wane, and they return to exist in the category of Susceptibles. No vaccination is considered here. This model is governed by the following system of ODEs:
\begin{equation}
    \begin{aligned}
        \dfrac{dS}{dt} &= \mu N -\beta \dfrac{SI}{N} + \zeta R  - \nu S \\
        \dfrac{dE}{dt} &= \beta \dfrac{SI}{N} -  \sigma E - \nu E \\
        \dfrac{dI}{dt} &= \sigma E  - \gamma I - \mu I \\
        \dfrac{dR}{dt} &= \gamma I - \nu R - \zeta R 
    \end{aligned} \qquad \qquad \text{subject to} \qquad \qquad
    \begin{aligned}
        S(t_0) &= S_0 \\
        E(t_0) &= E_0 \\
        I(t_0) &= I_0 \\
        R(t_0) &= R_0 \\
    \end{aligned}
\end{equation}
where $N = S+E+I+R$ is the total population, $\mu$ is the natural deaths rate, $\nu$ is the new births rate, $\beta$ is the infectious rate, $\sigma$ is the rate at which an Exposed person becomes Infectious, $\zeta$ is the rate which Recovered individuals return to the Susceptible statue due to loss of immunity, and $\gamma$ is the recovery rate.\\
According to the TFC framework, the latent solutions are approximated with the constrained expressions. That is,

\begin{equation}
    \begin{aligned}
      S &=\left(\mathbf{h} - \beta_1\mathbf{h_0} \right)\T \boldsymbol{\xi}_1 + \beta_1 S_0 \\
      E &=\left(\mathbf{h} - \beta_1\mathbf{h_0} \right)\T \boldsymbol{\xi}_2 + \beta_1 E_0 \\
      I &=\left(\mathbf{h} - \beta_1\mathbf{h_0} \right)\T \boldsymbol{\xi}_3 + \beta_1 I_0 \\
      R &=\left(\mathbf{h} - \beta_1\mathbf{h_0} \right)\T \boldsymbol{\xi}_4 + \beta_1 R_0 \\
    \end{aligned}
\end{equation}
The first four loss functions we present take into account the regression over the data. The last four losses drive the NN informing it with the physics governing the problem. The Loss functions are report below:
\begin{equation}
    \begin{aligned}
        \mathcal{L}_1 &= \Tilde{S} - S  \\
        \mathcal{L}_2 &= \Tilde{E} - E  \\
        \mathcal{L}_3 &= \Tilde{I} - I  \\
        \mathcal{L}_4 &= \Tilde{R} - R  \\
        \mathcal{L}_5 &= \Dot{S} - \mu N + \beta \dfrac{SI}{N} - \zeta R  + \nu S     \\
        \mathcal{L}_6 &= \Dot{E} - \beta \dfrac{SI}{N} +  (\sigma  + \nu ) E   \\
        \mathcal{L}_7 &= \Dot{I} - \sigma E  + (\gamma  + \mu) I \\
        \mathcal{L}_8 &= \Dot{R} - \gamma I + ( \nu + \zeta) R
    \end{aligned}
\end{equation}
To construct the Jacobian matrix $\mathcal{J}$ we need to compute the derivatives of the Losses in respect of the $\boldsymbol{\xi}$ to compute the approximate solutions of the state variables, whereas the other derivatives are essential to estimates the parameters (in this case $\beta$, $\gamma$, and $\sigma$) appearing in the system of eqs. \eqref{sir}. The resultant Jacobian matrix has the following form:
\begin{equation}
    \mathcal{J} = 
        \begin{bmatrix}
            \dfrac{\partial \mathcal{L}_1}{\partial \boldsymbol{\xi}_1}   &     \mathbf{0}   &    \mathbf{0}   &    \mathbf{0}   &     0  &  0   &   0  &   0\\    
            \mathbf{0}    &     \dfrac{\partial \mathcal{L}_2}{\partial \boldsymbol{\xi}_2}   &      \mathbf{0}    &    \mathbf{0}    &     0  &   0   &   0  &   0\\   
            \mathbf{0}  &    \mathbf{0}   &     \dfrac{\partial \mathcal{L}_3}{\partial \boldsymbol{\xi}_3}   &    \mathbf{0}   &     0  &   0    &   0   &   0\\    
            \mathbf{0}   &    \mathbf{0}   &     \mathbf{0}   &     \dfrac{\partial \mathcal{L}_4}{\partial \boldsymbol{\xi}_4}   &     0   &   0   &   0    &   0 \\    
            \dfrac{\partial \mathcal{L}_5}{\partial \boldsymbol{\xi}_1}   &     \dfrac{\partial \mathcal{L}_5}{\partial \boldsymbol{\xi}_2}   &     \dfrac{\partial \mathcal{L}_5}{\partial \boldsymbol{\xi}_3}   &     \dfrac{\partial \mathcal{L}_5}{\partial \boldsymbol{\xi}_4}   &   \dfrac{\partial \mathcal{L}_5}{\partial \beta}  &  0  & 0   &  \dfrac{\partial \mathcal{L}_5}{\partial \zeta} \\
            \dfrac{\partial \mathcal{L}_6}{\partial \boldsymbol{\xi}_1}   &     \dfrac{\partial \mathcal{L}_6}{\partial \boldsymbol{\xi}_2}   &     \dfrac{\partial \mathcal{L}_6}{\partial \boldsymbol{\xi}_3}   &     \dfrac{\partial \mathcal{L}_6}{\partial \boldsymbol{\xi}_4}  &    \dfrac{\partial \mathcal{L}_6}{\partial \beta}  &  0  &   \dfrac{\partial \mathcal{L}_6}{\partial \sigma} & 0 \\
            \mathbf{0}   &     \dfrac{\partial \mathcal{L}_7}{\partial \boldsymbol{\xi}_2}   &     \dfrac{\partial \mathcal{L}_7}{\partial \boldsymbol{\xi}_3}   &     \mathbf{0}  &     0  &  \dfrac{\partial \mathcal{L}_7}{\partial \gamma}   &  \dfrac{\partial \mathcal{L}_7}{\partial \sigma} & 0\\
           \mathbf{0}   &     \mathbf{0}   &     \dfrac{\partial \mathcal{L}_8}{\partial \boldsymbol{\xi}_3}   &     \dfrac{\partial \mathcal{L}_8}{\partial \boldsymbol{\xi}_4}  &     0  &  \dfrac{\partial \mathcal{L}_8}{\partial \gamma}   &  0 & \dfrac{\partial \mathcal{L}_9}{\partial \zeta}
        \end{bmatrix}
\end{equation}


\section{Results and Discussion}

\begin{figure}
\begin{subfigure}[h]{0.5\linewidth}
\includegraphics[width=\linewidth]{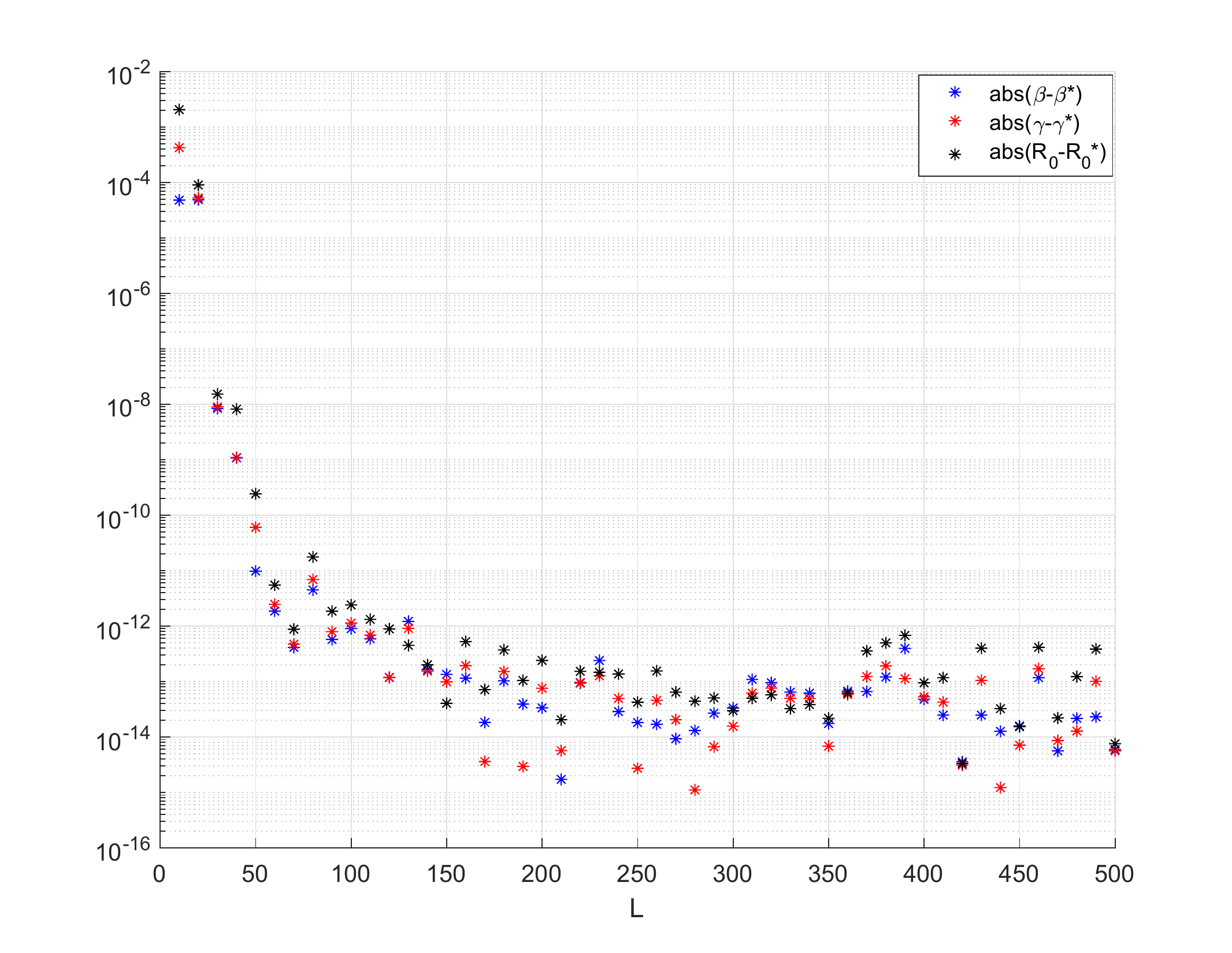}
\caption{$\delta=0$, $n=100$, and $L$ varies.}
\label{mc1_sir_l}
\end{subfigure}
\hfill
\begin{subfigure}[h]{0.5\linewidth}
\includegraphics[width=\linewidth]{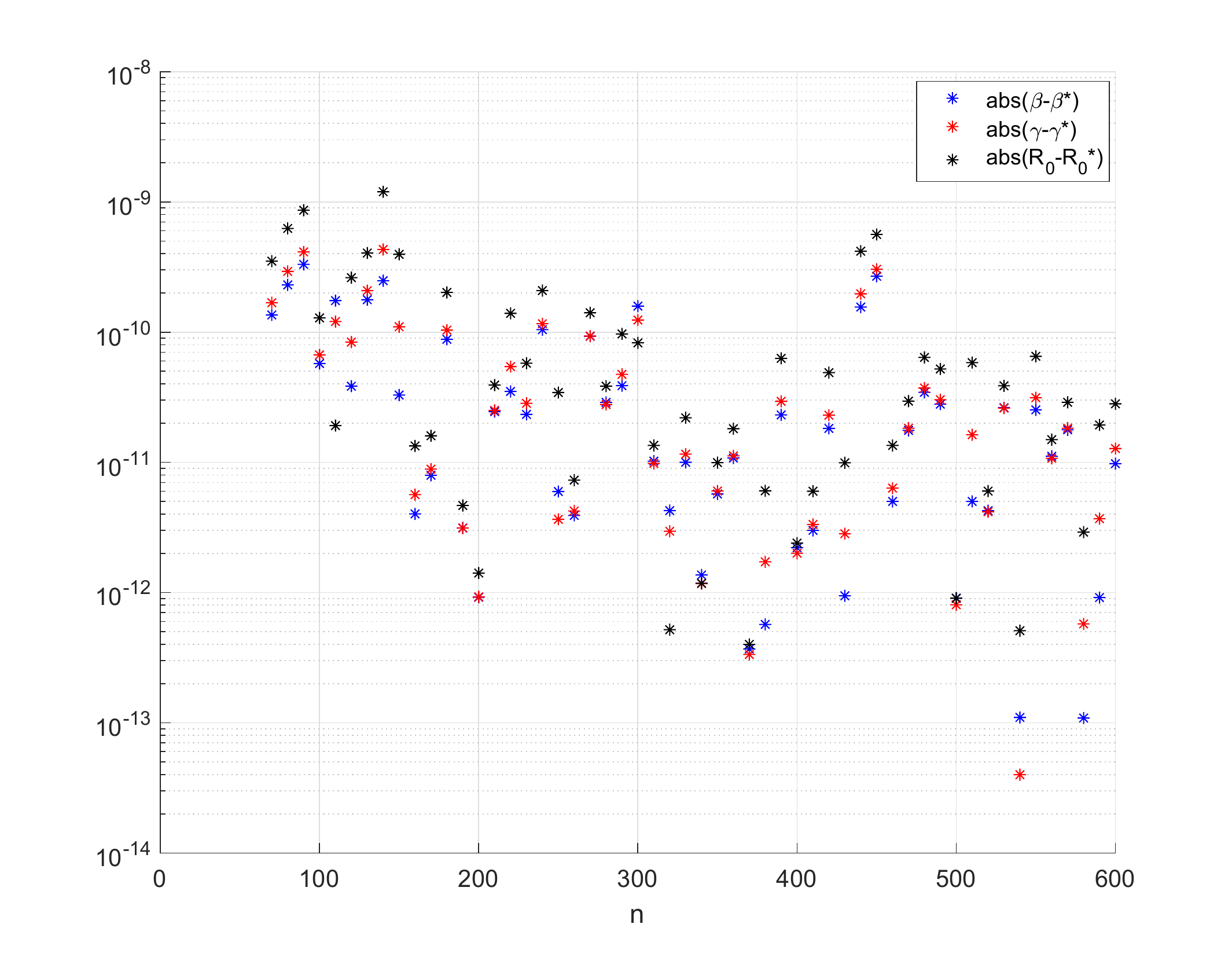}
\caption{$\delta=0$, $L=100$, and $n$ varies.}
\label{mc1_sir_n}
\end{subfigure}
\begin{subfigure}[h]{0.5\linewidth}
\includegraphics[width=\linewidth]{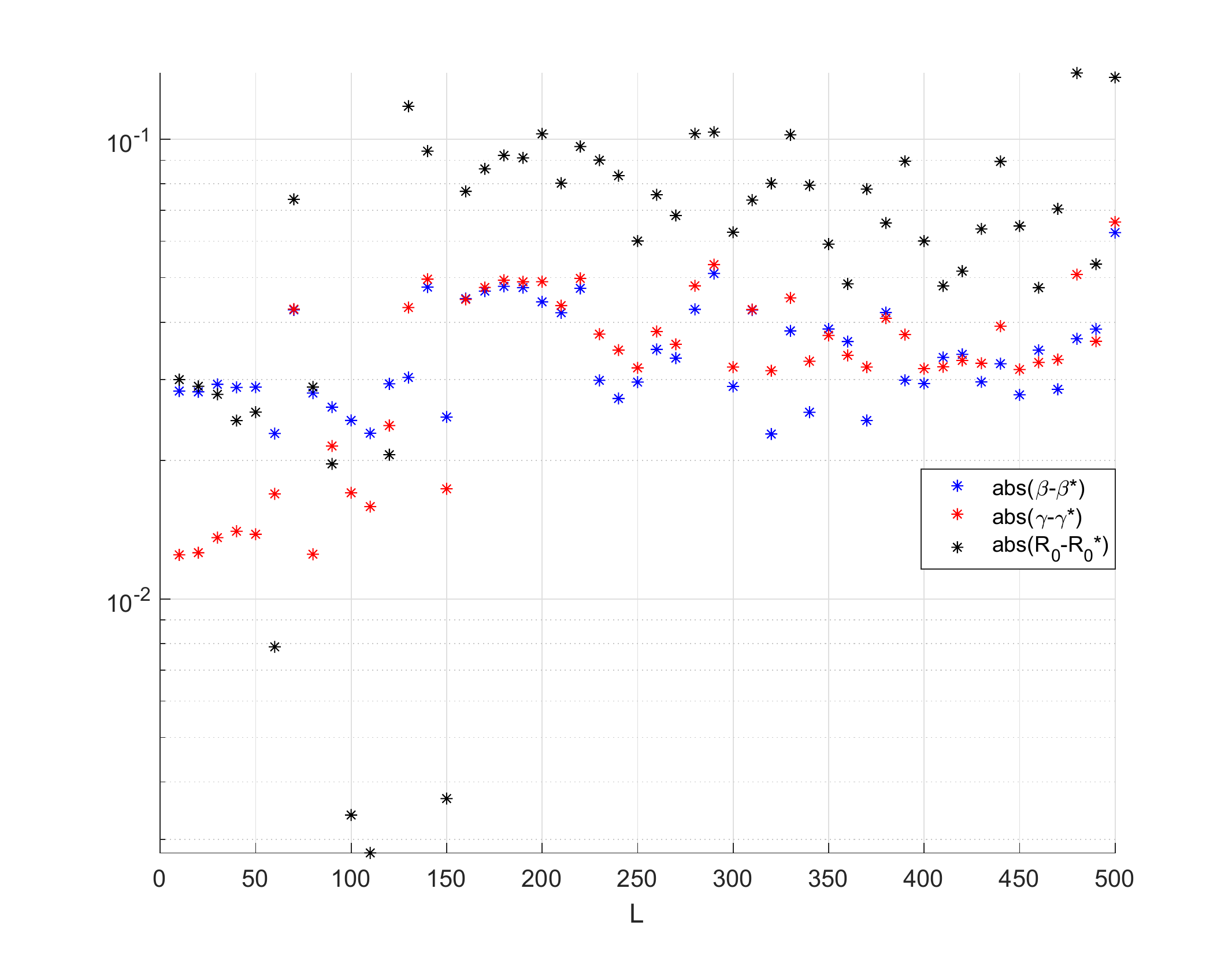}
\caption{$\delta=1$, $n=100$, and $L$ varies.}
\label{mc2_sir_l}
\end{subfigure}
\hfill
\begin{subfigure}[h]{0.5\linewidth}
\includegraphics[width=\linewidth]{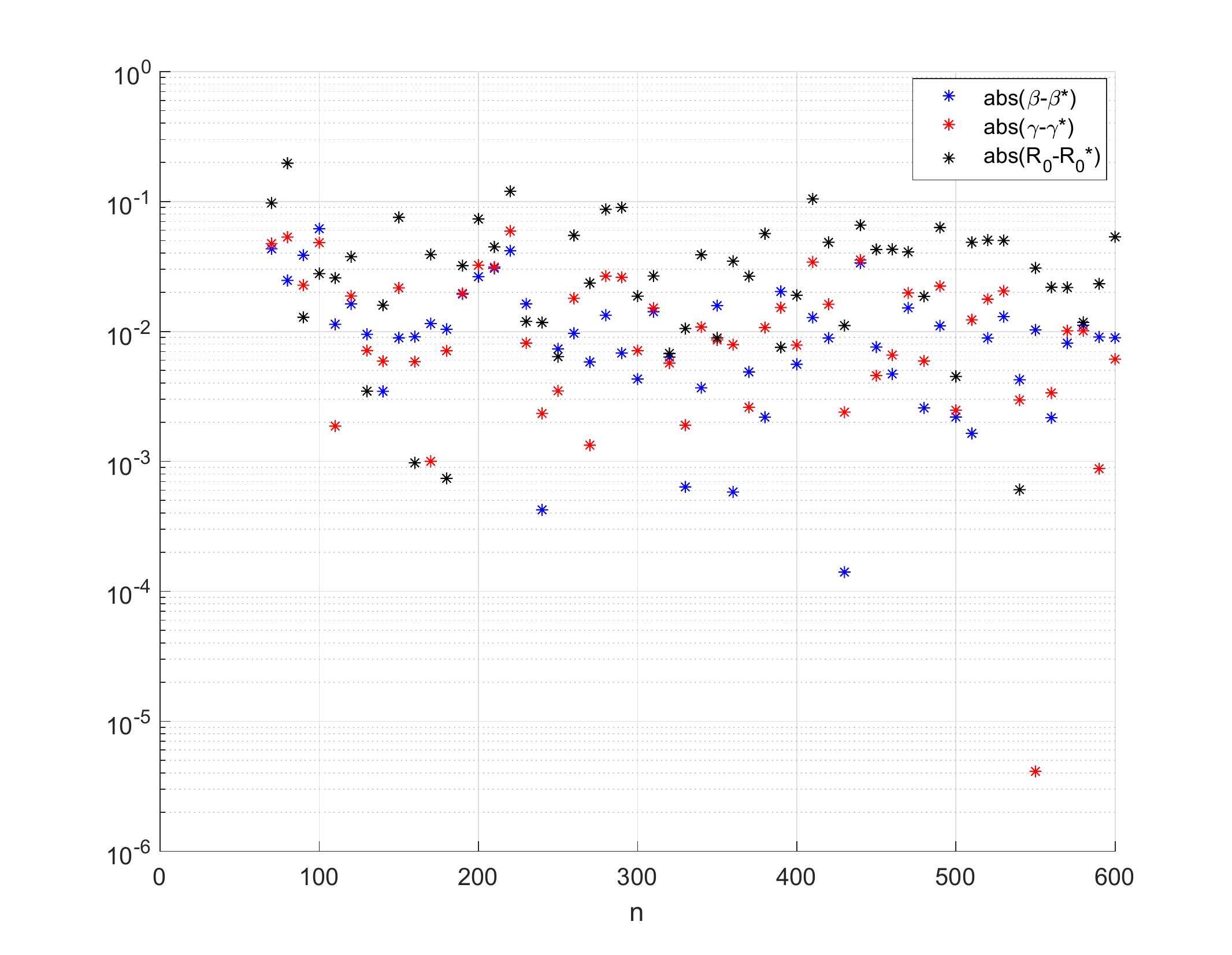}
\caption{$\delta=1$, $L=100$, and $n$ varies.}
\label{mc2_sir_n}
\end{subfigure}%
\caption{Monte Carlo simulations for SIR model with an \textit{ArcTan} activation function.}
\end{figure}

To test the ability of the X-TFC in performing data-driven parameters discovery of epidemiological compartmental models, we have created synthetic data-sets according to the three models presented above (SIR, SEIR, and SEIRS). In particular, for each model, a no-noisy synthetic data-set (here called original data-set $\Tilde{f}_{orig}$) has been generated by simply propagating the dynamics equations of the model using the MatLab function ODE113. In addition, to simulate a more realistic example, perturbed synthetic data-sets ($\Tilde{f}_{pert}$) have been created by adding noise to the original data-set. That is,
\begin{equation}
    \Tilde{f}_{pert} = \Tilde{f}_{orig} + \delta \, \text{unif}[-1,-1]
\end{equation}
where $\delta$ is the perturbation coefficient (equal to 0 for the original dataset) and $\text{unif}[\cdot,\cdot]$ represents a uniform distribution. The real values of the parameters governing the synthetic dataset are known, so that the accuracy of the results is measured by the absolute error between the real and estimated values of the parameters.\\
Additionally, the X-TFC method involves several hyperparameters that can be modified to obtain accurate solutions. These hyperparameters are the number of training points, $n$, the number of neurons, $L$, the type of activation function, and the probability distribution where input weights and bias are sampled from. Therefore, a sensitivity analysis has been performed to study the behaviour of the X-TFC method as these hyperparameters vary. The sensitivity analysis is only shown for the SIR model with no-noisy data, as a similar behaviour has been encountered for all the other models considered. First of all, the sensitivity analysis has demonstrated that, for the models analyzed, the solution accuracy is not as sensitive to the type of activation function used or to the probability distribution used to sample the inputs weights and bias as it is to the number of training points and the number of neurons, confirming the results found from the sensitivity analysis reported in \cite{etfc}. Hence, the two parameters that strongly influence the performances of the X-TFC are $n$ and $L$.
Figures \ref{mc1_sir_l} and \ref{mc1_sir_n} refer to the analysis with the original datatset ($\delta=0$). As illustrated, high values of $L$ ($L>150$), with a fixed $n$, do not lead to an improvement of the accuracy, since Fig.\ref{mc1_sir_l} presents an asymptotic-like behaviour. The same considerations are valid varying $n$ and keeping fixed $L$ (Fig.\ref{mc1_sir_n}). Indeed, the solution does not significantly improves increasing the number of discretization points. This result is obtained also if a perturbed dateset is considered (see Fig.\ref{mc2_sir_n}). On the other hand, Fig. \ref{mc2_sir_l} shows an interesting behaviour. The accuracy of the solution gets worse by increasing the number of neurons $L$. This trend is probably due to the fact that X-TFC tries to overfit the perturbed data, so that to diverge too much from the real curves and thus obtaining a not accurate estimation of the parameters. 
The rest of this section focuses on the results obtained for each model presented previously. For these problems, the \textit{ArcTan} activation function and a uniform random distribution ranging within [-10,10] are employed for the ELM.

\subsection{SIR Model}

\begin{figure}
\begin{subfigure}[h]{0.5\linewidth}
\includegraphics[width=\linewidth]{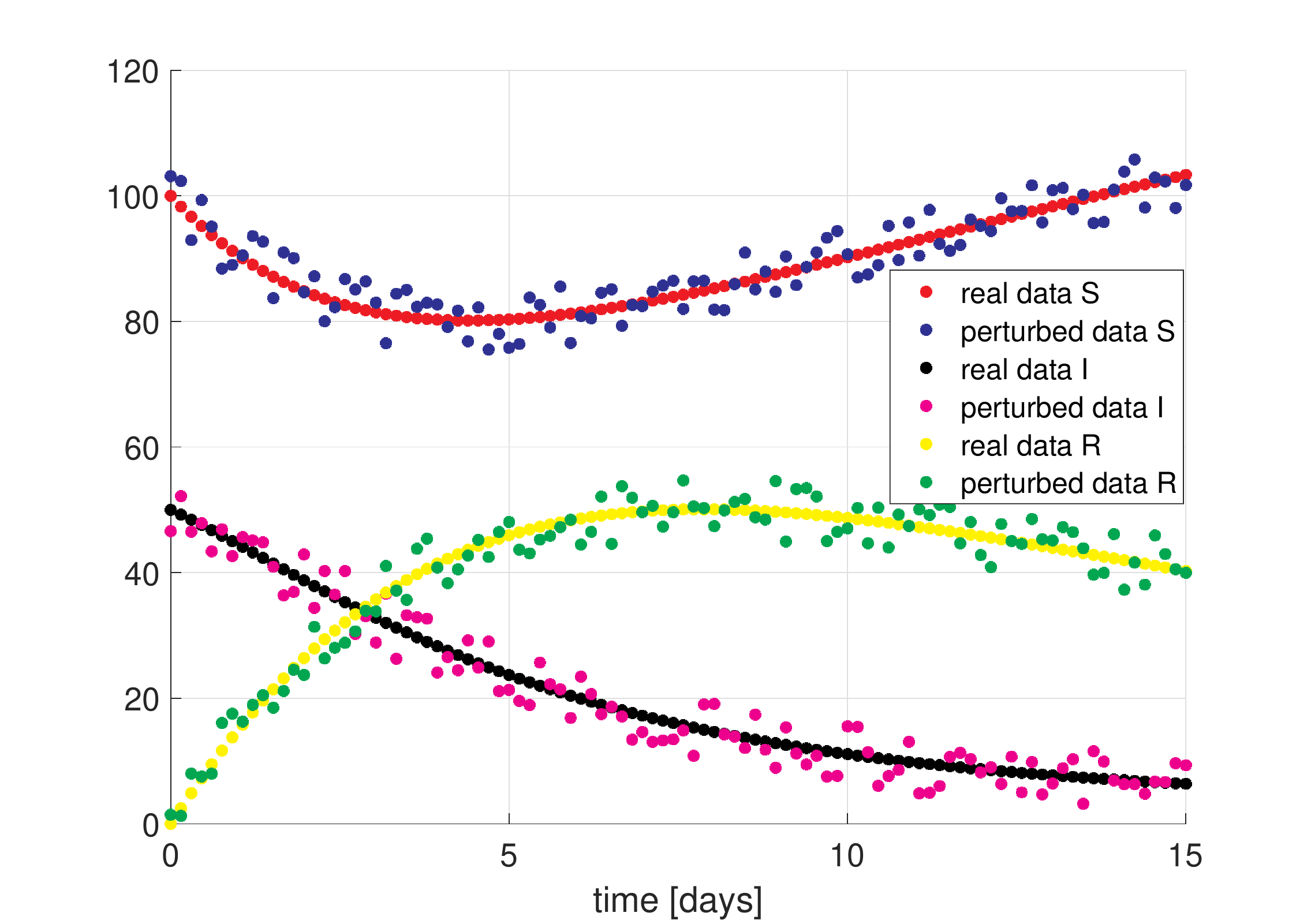}
\caption{Real and perturbed data.}
\label{sir_data}
\end{subfigure}
\hfill
\begin{subfigure}[h]{0.5\linewidth}
\includegraphics[width=\linewidth]{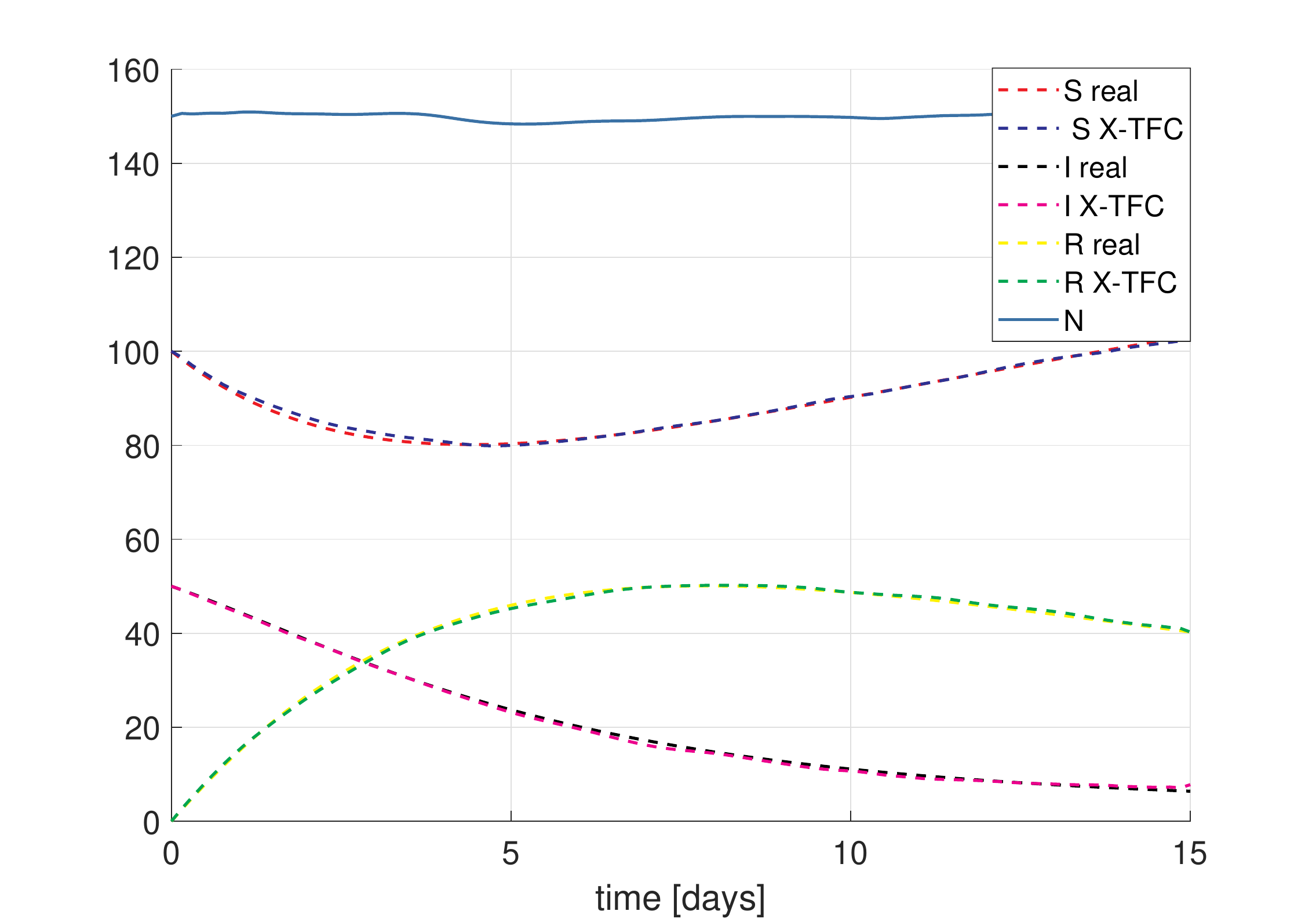}
\caption{Results with perturbed data.}
\label{sir_res}
\end{subfigure}
\caption{Results for SIR model with $\delta=5$, $n=100$, and $L=50$.}
\end{figure}

Here, the results and the performances for SIR problem are shown. The outputs are obtained by setting the following parameters:\\
\begin{itemize}
    \item natural mortality rate: $\mu = 0.1$ (set equal to the birth rate, to simulate a constant number of population); 
    \item effective contact rate (possibility to be infected): $\beta=\frac{1}{2}$; 
    \item removal rate (how often infected people become recovered): $\gamma=\frac{1}{3}$;
    \item initial conditions: $ S_0 = 100$; $I_0 = 5$; $R_0 = 0$;
    \item analysis time: 15 days.
\end{itemize}
Several simulations are carried out by varying the intensity of the noise, and the outputs are reported in Table \ref{tab:sir}. While we could find the exact values of parameters with the original dataset, a slight deviation of these values occurs by increasing the perturbation coefficient $\delta$. However, the absolute errors on the parameters result to have at least 2 digits of accuracy. Figures \ref{sir_data} and \ref{sir_res} report the perturbed and real datasets and the solution of the problem for the case of $\delta=5$, respectively. As it can be seen, the X-TFC is able to obtain an accurate solution avoiding the overfitting on the data, as it could be expected by a simple regression on the perturbed dataset. This is due to the information about the physics of the problem, which acts as a regulator, that are embedded in the physics-informed training framework. The accuracy of the inversion with perturbed datasets is also proved by the constant value of the population $N$, as it has to be from the theory.

\begin{table}[H] 
\caption{Performances of the proposed physics-informed framework in the data-driven discovery of the SIR model with different noise on the data, with $n= 100$ and $L= 50$.}

\footnotesize
\label{tab:sir}
    \begin{center}
    \begin{tabular}{ccccccccc} 
             \hline
              Noise & $\#$ of iterations & CPU time [s] & $\beta$ & $|\text{err}(\beta)|$ & $\gamma$ & $|\text{err}(\gamma)|$ & $\mathcal{R}_0$ & $|\text{err}(\mathcal{R}_0)|$ \\
             \hline
              0  & 2 & 0.002 & 0.5000 & 0 & 0.3333 & 0 & 1.500 &  0\\
            0.1  & 4 & 0.036 & 0.4999 & $4.2\times 10^{-5}$ & 0.3334 & $4.2\times 10^{-5}$ & 1.4997 & $3.2\times 10^{-4}$ \\
              1  & 7 & 0.049 & 0.4996 & $4.4\times 10^{-4}$ & 0.3338  &$4.2\times 10^{-4}$  & 1.4968 & $3.2\times 10^{-3}$ \\
             5  & 8 & 0.051 & 0.4978 & $2.2\times 10^{-3}$ & 0.3353 & $2.1\times 10^{-3}$ & 1.4884 & $1.56\times 10^{-2}$ \\
            \hline
    \end{tabular}
    \end{center}
\end{table}

\subsection{SEIR Model}

Here, the results and the performances for SEIR problem are shown. The outputs were obtained by setting the following parameters:\\
\begin{itemize}
    \item natural mortality rate $\mu = 0.5$ (set equal to the birth rate, to simulate a constant number of population); 
    \item vaccine rate $\nu = 0.5$;
    \item effective contact rate (possibility to be infected) $\beta=0.3$; 
    \item removal rate (how often infected people become recovered) $\gamma=0.6$;
    \item progression rate from exposed to infected $\sigma=0.9$;
    \item initial conditions: $ S_0=70$; $E_0=30$; $I_0=10$; $R_0=0$;
    \item days = 15.
\end{itemize}

\begin{figure}
\begin{subfigure}[h]{0.5\linewidth}
\includegraphics[width=\linewidth]{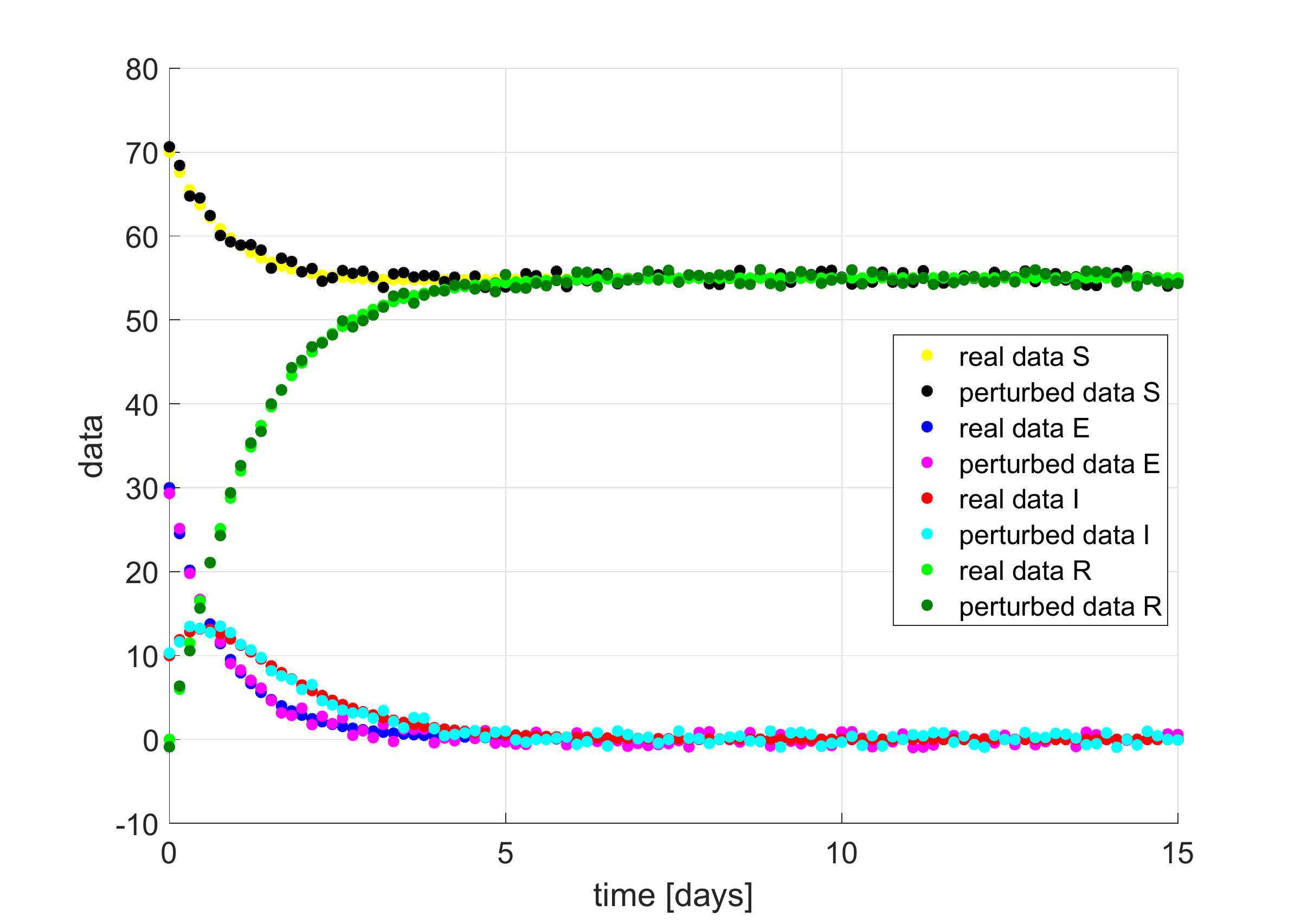}
\caption{Real and perturbed data.}
\label{fig:seir_data}
\end{subfigure}
\hfill
\begin{subfigure}[h]{0.5\linewidth}
\includegraphics[width=\linewidth]{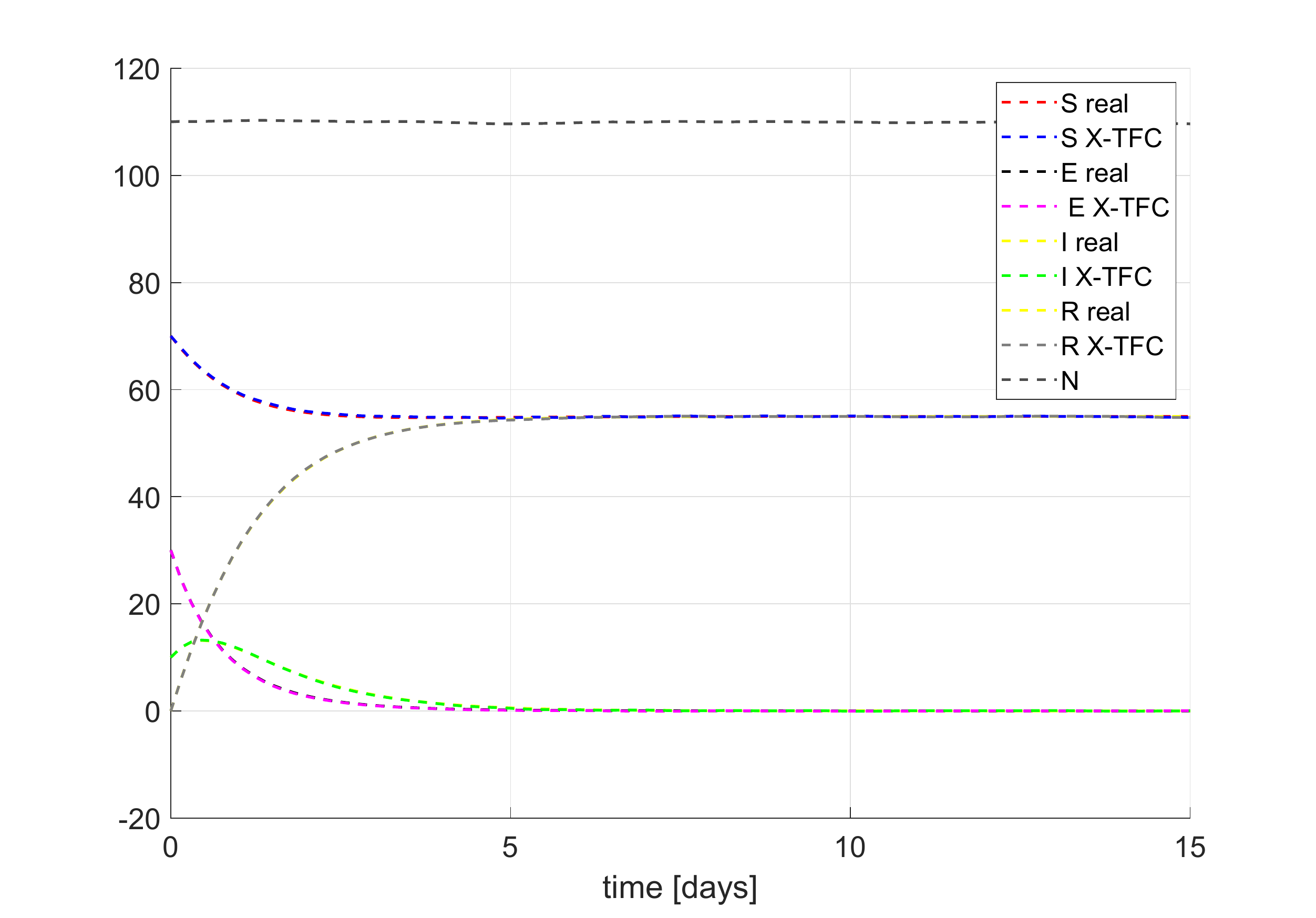}
\caption{Results with perturbed data.}
\label{fig:seir_results}
\end{subfigure}
\caption{Results for SEIR model with $\delta=3$, $n=100$, and $L=80$.}
\end{figure}

Several simulations are carried out by varying the intensity of the noise, and the outputs are reported in Table \ref{tab:seir}. While we could find the exact values of parameters with the original dataset, a slight deviation of these values occurs by increasing the perturbation coefficient $\delta$. However, the absolute errors on the parameters result to have at least one digit of accuracy. Figures \ref{fig:seir_data} and \ref{fig:seir_results} report the perturbed and real datasets and the solution of the problem for the case of $\delta=3$, respectively. Again, the X-TFC is able to obtain an accurate solution avoiding the overfitting on the data, as it could be expected by a simple regression on the perturbed dataset.

\begin{table}[H] 
\caption{Performances of the proposed physics-informed framework in the data-driven discovery of the SEIR model with different noise on the data, with $n= 100$ and $L= 80$, $\mu = \nu =0.5$}
\footnotesize
\label{tab:seir}
    \begin{center}
    \begin{adjustbox}{max width=\textwidth}
    \begin{tabular}{ccccccccccccc} 
             \hline
              Noise & $\#$ of iterations & CPU time [s] & $\beta$ & $|\text{err}(\beta)|$ & $\gamma$ & $|\text{err}(\gamma)|$ &  $\sigma$ & $|\text{err}(\sigma)|$ & $\mathcal{R}_0$ & $|\text{err}(\mathcal{R}_0)|$ \\
             \hline
              0   & 3  & 0.004 & 0.3 & 0 & 0.6 & 0 & 0.9 & 0  & 0.5 & 0 \\
              0.1 & 3 & 0.051 & 0.2971 & $2.9 \times 10^{-3}$ & 0.5996 & $3.7 \times 10^{-4}$ & 0.9005 & $4.9 \times 10^{-4}$ & 0.4955 & $4.5 \times 10^{-3}$ \\
               1  & 20 & 0.27 & 0.2711 & $2.9 \times 10^{-2}$ &0.5962 &  $3.8 \times 10^{-3}$ & 0.9048 &  $4.8 \times 10^{-3}$ & 0.4547 & $4.5 \times 10^{-2}$  \\
               3  & 45 & 0.61 & 0.2130 & $8.7 \times 10^{-2}$ & 0.5878&  $1.2 \times 10^{-2}$ & 0.9141 &  $1.4 \times 10^{-2}$ & 0.3624 & $1.4 \times 10^{-1}$  \\
            \hline
    \end{tabular}
    \end{adjustbox}
    \end{center}
\end{table}

\subsection{SEIRS Model}

Here, the results and the performances for SEIRS problem are shown. The outputs were obtained by setting the following parameters:\\
\begin{itemize}
    \item natural mortality rate $\mu = 0.5$ (set equal to the birth rate, to simulate a constant number of population); 
    \item effective contact rate (possibility to be infected) $\beta=0.3$; 
    \item removal rate (how often infected people become recovered) $\gamma=0.6$;
    \item progression rate from exposed to infected $\sigma=0.9$;
    \item rate which recovered individuals return to the susceptible statue (due to loss of immunity) $\zeta=0.5$;
    \item initial conditions: $ S_0=70$; $E_0=30$; $I_0=10$; $R_0=0$;
    \item days = 15.
\end{itemize}

\begin{figure}
\begin{subfigure}[h]{0.5\linewidth}
\includegraphics[width=\linewidth]{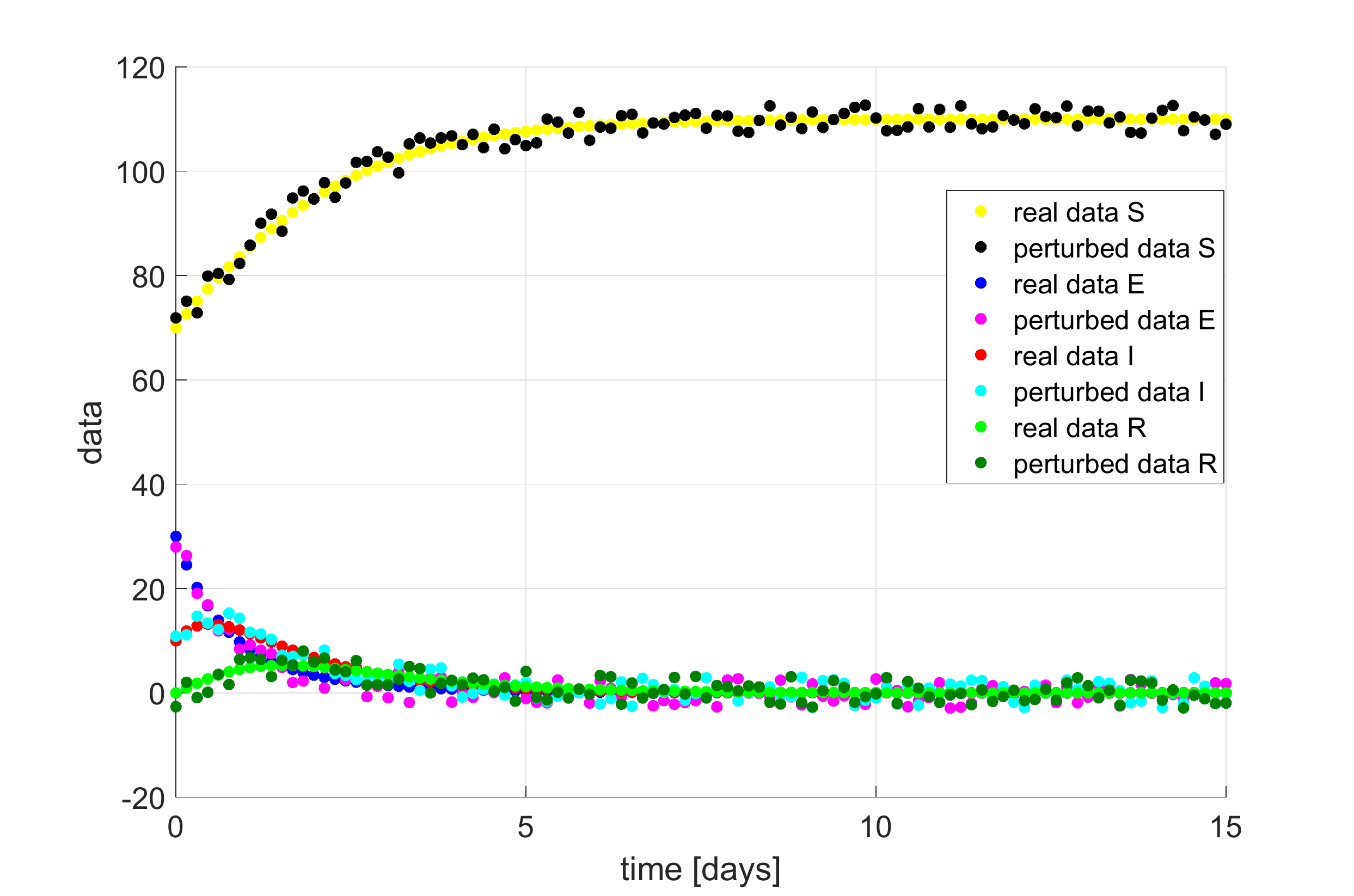}
\caption{Real and perturbed data.}
\label{fig:seirs_data}
\end{subfigure}
\hfill
\begin{subfigure}[h]{0.5\linewidth}
\includegraphics[width=\linewidth]{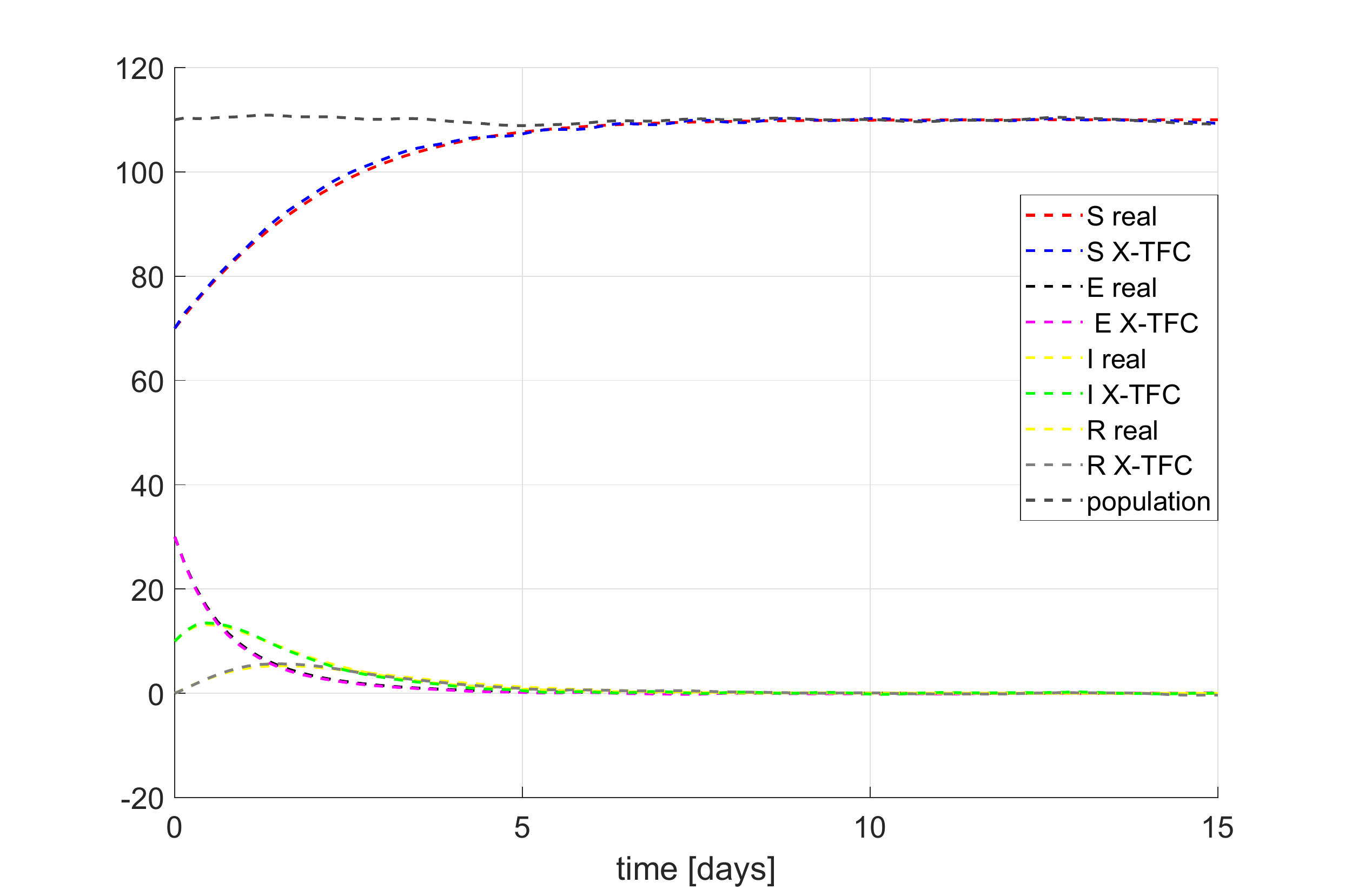}
\caption{Results with perturbed data.}
\label{fig:seirs_results}
\end{subfigure}
\caption{Results for SEIRS model with $\delta=3$, $n=100$, and $L=80$.}
\end{figure}

Several simulations are carried out by varying the intensity of the noise, and the outputs are reported in Table \ref{tab:seirs}. While we could find the exact values of parameters with the original dataset, a slight deviation of these values occurs by increasing the perturbation coefficient $\delta$. However, the absolute errors on the parameters result to have at least two digits of accuracy. Figures \ref{fig:seirs_data} and \ref{fig:seirs_results} report the perturbed and real datasets and the solution of the problem for the case of $\delta=3$, respectively. Again, the X-TFC is able to obtain an accurate solution avoiding the overfitting on the data, as it could be expected by a simple regression on the perturbed dataset.

\begin{table}[H] 
\caption{Performances of the proposed physics-informed framework in the data-driven discovery of the SEIRS model with different noise on the data, with $n=100$ and $L= 80$, $\mu = 0.5$}
\footnotesize
\label{tab:seirs}
    \begin{center}
    \begin{adjustbox}{max width=\textwidth}
    \begin{tabular}{ccccccccccccc} 
             \hline
              Noise & $\#$ of iterations & CPU time [s] & $\beta$ & $|\text{err}(\beta)|$ & $\gamma$ & $|\text{err}(\gamma)|$ &  $\sigma$ & $|\text{err}(\sigma)|$ &  $\zeta$ & $|\text{err}(\zeta)|$ & $\mathcal{R}_0$ & $|\text{err}(\mathcal{R}_0)|$ \\
             \hline
              0   & 3  & 0.04 & 0.3 & 0 & 0.6 & 0 & 0.9 & 0  & 0.5 & 0 & 0.5 & 0 \\
              0.1 & 14 & 0.2 & 0.3011 & $1.1 \times 10^{-3}$ & 0.6020 & $2.0 \times 10^{-3}$ & 0.9026 & $2.6 \times 10^{-3}$ & 0.5028 & $2.8 \times 10^{-3}$ & 0.500 & $1.7 \times 10^{-4}$  \\
              1  & 5 & 0.09 & 0.3093 & $9.3 \times 10^{-3}$ &0.6183 &  $1.8 \times 10^{-2}$ & 0.9249 &  $2.5 \times 10^{-2}$ & 0.5251 & $2.5 \times 10^{-2}$ & 0.5002 & $2.1 \times 10^{-4}$  \\
              3  & 110 & 1.36 & 0.3174 & $1.7 \times 10^{-2}$ & 0.6465&  $4.7 \times 10^{-2}$ & 0.9680 &  $6.8 \times 10^{-2}$ & 0.5576 & $5.8 \times 10^{-2}$ & 0.4909 & $9.1 \times 10^{-3}$  \\
            \hline
    \end{tabular}
    \end{adjustbox}
    \end{center}
\end{table}

\section{Conclusions and Outlooks}

In this work, the new physics-informed framework X-TFC has been employed to solve data driven discovery of DEs, also called inverse problems, via a deterministic approach. In particular, compartmental epidemiological models (SIR, SEIR, SEIRS) have been taken into account as test problems. The goal was to retrieve the parameters governing the dynamics equations considering unperturbed and perturbed data, to better simulate the reality. The tests have shown very good results even when a significant noise was added to the data. Furthermore, the information about the physics of the problem (considered for the training of the X-TFC) has allowed to avoid the over-fitting and thus to obtain good estimations of parameters with noisy data.
The low computational times obtained are extremely important to process data as soon as they are acquired, so that the results can be updated in real-time. Moreover, the good estimations of parameters allow to make predictions about the imminent future: this makes it possible to take actions in the short term (as it should be in emergency scenarios). Future works involve the inversion of models with non-constant parameters (i.e. parameters that follow mathematical laws) as well as probabilistic parameters estimation (via Bayesian fashion) in different research fields, such us business, biology, space and nuclear engineering.\\

\section*{Conflicts of Interest}
The authors declare no conflict of interest. 
\bibliographystyle{ieeetr}
\bibliography{Refs}

\end{document}